\documentclass[aps,twocolumn,amssymb,showpacs,amsmath,preprintnumbers]{revtex4}
\usepackage{graphicx}
\usepackage{bm}

\begin{document}
\title{
Phase heterogeneities of lipidic aggregates}

\author{L. V.~Elnikova}
\affiliation{
A. I. Alikhanov Institute for Theoretical and Experimental Physics, \\
25, B. Cheremushkinskaya st., 117218 Moscow, Russia}
\date{\today}

\begin{abstract}
We propose a model for explanation the "domain-wall" type configuration states in binary lipid mixtures of cationic and neutral lipids, associated with observed relaxation effects in their aggregates. We apply the analogy with formation of Kibble-\.Zurek topological defects, which we suppose connected with structural dynamics of the lipid phases. In frames of the proposed model, the density of kink-type defects and the energy of the configurations are calculated.
\end{abstract}

\maketitle

\section{Introduction}
A number of applications for lipids and their mixtures in the biomaterial technology, therapy and industry cause the necessity of careful theoretical predictions of their phase transformations. Lipidic aggregates, as well as lyotropic systems at all, are ideal substances demonstrating the quantum topological phase transitions
(\cite{Kleman, El0601651} and references therein). Some experiments on polar fluids~\cite{Wyss, epl, Castellano, Kiselev1, To} reveal
new interesting phenomena of self-organization.

For concentrated suspensions, pastes, emulsions, foams, and associative polymers,
the mechanism of structure relaxation in soft solids, which is based on the mechanical
anelastic spectroscopy in rheological frequencies, was proposed by Wyss and coworkers \cite{Wyss}.
Amplitude-dependent measurements have shown, that when the strain rate becomes
large, it can itself drive the slow structural relaxation process at the time
scale of the imposed strain rate \cite{Wyss}. Also, under an applied strain, the
observations of high-frequency shifts cause to interpret the enigmatical
slow relaxation dynamics in a principally new way. While, the authors \cite{Wyss} seek an explanation of their measurements
in analogous phenomena in supercooled fluids \cite{epl}.

From the other side, the recent anelastic spectroscopy studies
\cite{Castellano} of dynamical processes on neutral DMPC
(dimyristoil phosphatidylcholine), DOPE (di-oleoyl
phosphotiylethanolamine) lipids, cationic DDAB
(dimethyldioctadecylammonium), DOTAP (di-oleoyl trimethylammonium)
lipids, and their DOTAP/DOPE and DOTAP/DMPC mixtures
brought out the hypothesis of  new micro- and nanoscale structure heterogeneities
in the lipid membranes \cite{Castellano}, which look like onto domain walls. These observations have been carried out in a wide
temperature range starting with the supercooled state,
and at wide-range excitation frequencies ($10^2$--$10^4$ Hz), with the lipids deposited on a solid substrate.

Weak frequency-dependent shifts in the elastic modulus
are observed at these cryogenic temperatures; peaks on
the relaxation curves at the low-dynamics regime are
closed by the typical "smooth" relaxation,
so, the observations evidence collective short-range motions of the lipids \cite{Castellano}.

Also, one may indirectly compare these data with neutron scattering
\cite{Kiselev1} and atomic-force microscopy (AFM) \cite{To} on some neutral
and cationic lipids and their mixtures.

Hence, a number of soft matter phenomena can be described in
the common interdisciplinary modeling associated with a mechanism
of forming of topological defects, such as domain walls, or vortex strings.

\section{Formation of topological defects }
Generally, in the cases specified above, the phase transitions
are continuous. The Kibble-\.Zurek (KZ) mechanism of formation of
the domain-wall type defects in an adiabatic regime is convenient
for a description of dynamics of these transitions \cite{Zurek,
ActaPol}.

As topological defects, domain walls are broadly known in their
universality \cite{Seidel, 0504148}, from cosmological models to the theory of
condensed matter \cite{Kleman, 1832}. The initial Landau-Zener (LZ) Hamiltonian \cite{Zener} of a two-level
system expresses the dynamical processes at classical phase transitions.
Afterward, the LZ Hamiltonian has been generalized onto quantum phase transitions.	

Here, we identify the domain-wall defect formation with the
some quantum phase transition at the last ("adiabatic")
stage of the LZ-type evolution, which includes three regimes in general \cite{Zurek}. Then, one may apply a formalism of adiabatic quantum computations,
or Quantum Annealing (AQC-QA), with the KZ approach \cite{1832} in absence of frustrations.

This modeling enables us to estimate a density of kinks and a residual energy,
corresponding to the one-dimensional quantum Ising system with
the time-dependent term of a transverse field, the Hamiltonian of which is~\cite{1832}:
\begin{equation}\label{1}
H(t)=-\sum_iJ_i\sigma_i^{z}\sigma_{i+1}^z-\Gamma(t)\sum_ih_i\sigma_i^x,
\end{equation}
here, $\sigma_i^{x,z}$ are Pauli matrices for the $i$-th spin of the chain. % of an $L$ size,
 $J_i$ denote random couplings between neighboring spins, and $h_i$ is
a random transverse field. The function of time $\Gamma(t)$ serves for rescaling a
transverse field $h_i$  at an annealing rate $\tau^{-1}$:
\begin{equation}
\Gamma(t)=-\frac{t}{\tau}, t\in(-\infty, 0].
\end{equation}
One should note, that following the KZ scenario, in the end of the ordering into a non-equilibrium state, the transition time $\tau_Q$ and the average finite ordered domain size are connected by
\begin{equation}
\hat{\xi}\simeq\tau_Q^{z/z(\nu+1)},
\end{equation}
where $z$ and $\nu$ are critical exponents \cite{Dz2006}.

The LZ Hamiltonian describes the evolution of a system in time $t$
\begin{equation}
\tilde{H}=\frac{1}{2}
(\begin{array}{cc}
\frac{t}{\tau_Q} &  1 \\
1 & -\frac{t}{\tau_Q}
\end{array}).
\end{equation}

In the adiabatic-impulse approach,
with the evolution of a system in time $t$ from $-\infty$
to 0, after completion of the transition, the density of kinks equals:
\begin{equation}
n=\lim_{N=+\infty}\langle\frac{1}{2N}\sum_{n=1}^{N-1}(1-\sigma_n^z\sigma_{n+1}^z)\rangle.
\end{equation}
In other words, dynamics in a system can be
exactly described by a series of uncoupled LZ systems \cite{15, pra063405}.

The authors of \cite{1832} have analyzed the fermionic Hamiltonian
received by applying the Jordan-Wigner transformation to the Hamiltonian (1). To treat the question of dynamics, they have solved a system
of linear differential equations received from the fermionic
Bogoliubov's equations by means of the known \textit{ansatz}.
So, in agreement with~\cite{Dz2006}, at a finite $\tau$ and at LZ factor, scaled by
Bogoliubov-de Gennes transformation, the density
of kinks and the residual energy, reduced to the total lattice
size $L$, can be estimated respectively by the next relations \cite{1832}:
\begin{equation}
\rho_k(\tau)\sim\frac{1}{\widetilde{L}_\epsilon(\tau)} \geq
\frac{[\Pi^{-1}(\epsilon)]^{2}}{\log^2(\gamma \tau)}
\end{equation}
\begin{equation}
[\frac{E_{res}}{L}] \sim \frac{1}{\log^\zeta(\gamma\tau)},
\end{equation}
The $\zeta$ parameter has been found numerically
$\approx3.4\pm0.2$~\cite{1832}. $\gamma$ are the Bogoliubov
operators diagonalizing $H(t)$, $\Pi$ is the universal function, and $\Pi^{-1}$ denotes the inverse function of
$\Pi$ \cite{1832}; the characteristics $g$ is defined by the equality
$g=-\log(-\triangle_1)/\sqrt{L}$,
$\triangle_1=2(\epsilon_1+\epsilon_2)$ is the excitation energy of
single-particle eigenvalues $\epsilon_1 \leq \epsilon_2 \leq ...
\epsilon_L$. $E_{res}=E_{t}-E_{classical}$, here $E_{t}$ denotes a time-evolved
state energy, and $\widetilde{L_\epsilon}(\tau)$ is a length of the
defect-free region upon annealing. The critical point probability
(\textit{ibid}) is
\begin{equation}
P^{cr.point}(\tau,L)\approx\Pi(g_c)\equiv\int_{g_c}^{\infty}dg P(g).
\end{equation}
$g_c$ denotes the characteristics g in the critical point. The exact chain's $P$ depending on concentration is known ~\cite{pra063405}.)

In principle, we know the classical 2D Ising simulation with the
Glauber dynamics ~\cite{ActaPol} (the heat bath algorithm) for a
non-equilibrium system under the K\.{Z} mechanism. The continual version of the Hamiltonian with pure
relaxation time is given there. It seems useful for our
goal, because, in such model, the "domain walls"
are always annihilating ~\cite{ActaPol}.
However, we can not follow it directly by virtue of the reasons shown below.

\section{Numerical modeling and results}
According to the hypothesis of displacing lipids motion ~\cite{Castellano}, we carry out the numerical experiments in the spirit of the quantum model
of ~\cite{1832} and references therein, but for a 3D Ising lattice allowing frustrations.

To involve "concentration" in this modeling, we have to
keep a number of particles during simulations.

At free field parameters, let us assume that the Hamiltonian (1)
is a bosonized Hamiltonian of our particles, so that we operate
with the spin variables $\sigma_{ij}=\pm1$.
\begin{figure}
\includegraphics*[width=75mm]{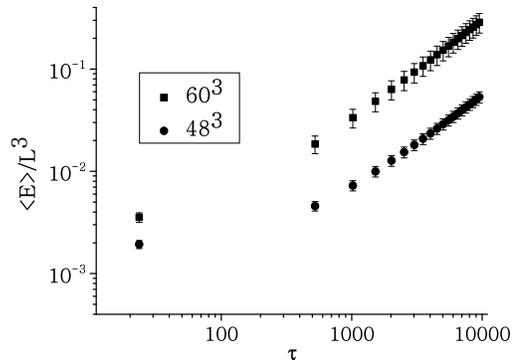}
\caption{\small Average residual energy per site as a function of the annealing rate $\tau$.}
\end{figure}
\begin{figure}
\includegraphics*[width=75mm]{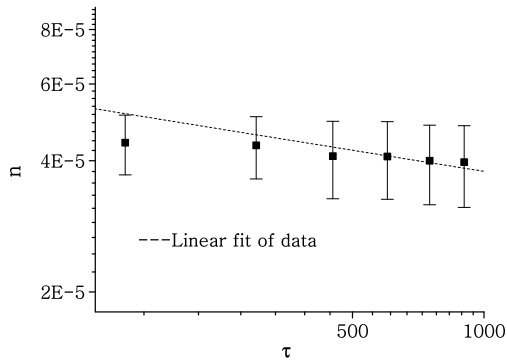}
\caption{\small Density of kinks as a function of the annealing rate
$\tau$ for the $60\times60\times60$ lattice.}
\end{figure}

Then, above the critical point~\cite{Kibble_2}, at the periodical boundary conditions on the $bcc$-lattice
of $48\times48\times48$ and $60\times60\times60$ sizes, we calculate this energy
(Fig.1) and the kink density (Fig. 2.).

For the Hamiltonian (1), we have performed a typical Monte Carlo (MC) algorithm,
in which the probability of states is $exp(-\triangle H/k_BT)$, where
 $T$ is absolute temperature, $k_B$ is the Boltzmann factor, and $\triangle H$ is
the energy difference between neighboring states. A number of MC
running steps equals $10^6$, and the thermalization is carried out in $10^5$ steps.

\section{Discussion}
At the adiabatic regime (Fig. 1), the average residual energy behaves in agreement with the K\.Z mechanism (see, for instance, \cite{1832}). These data allow us to discuss, how far the LZ theory is satisfied to the hypothesis, following from the experiment \cite{Castellano}, where a density of kinks could be measured hereafter. The lattice models for lipids are widely known \cite{latlip}; in their frames, an aggregation scale and a lattice size are comparable.

Thus, it will be possible to construct a structure parameter of fluid aggregates and lipidic mixtures in terms of quantum Ising models \cite{Zurek}.

Some interesting 3D calculations \cite{17}  are known for vortex strings in the $He$ isotopes, which qualitatively agree with our results (Fig. 2.).

However, our approach encloses the next problem. The K\.Z-type models contain a concentration dependence, an absence of which was emphasized especially for the cryogenic experiments \cite{Castellano} on the lipid mixtures. Sometimes \cite{15, pra063405}, at the calculations of domain wall sizes in the one-dimensional case, this question is imperceptible, but is not solved for different types of soft solids \cite{Wyss, Castellano}.

\section{Conclusion}
So, in contrast to lipidic phases without defects, where phase transformations may be characterized in classical Ising models \cite{20}, the case of unexplored low-temperature phase transitions compels us to involve the quantum lattice model with a random transverse field.

If the domain walls are annihilating and/or generating a new phase, then it is reasonable to continue the experiments in a wide amplitude and frequency range. The discussed model can be tested also in neutron diffraction experiments on similar lipidic membranes~\cite{Kiselev1}.

In the case of AFM, and in general in the presence of a substrate~\cite{To, substrate}, the observable relaxation peaks have to be separately specified.
$$
$$
The author thanks Prof. F. Tokumasu and Prof. R. Cantelli for
useful discussions, and also Prof. I. S. Golovin for his help
in data processing. For the simulations, the FORTRAN programs
composed jointly with Prof. V. A. Kashurnikov, were adapted.

\end{document}